%% file: templateArxiv.tex
\documentclass{article}

\usepackage{PRIMEarxiv}

\usepackage[utf8]{inputenc} 
\usepackage[T1]{fontenc}    
\usepackage{hyperref}       
\usepackage{url}            
\usepackage{booktabs}       
\usepackage{amsfonts}       
\usepackage{nicefrac}       
\usepackage{microtype}      
\usepackage{lipsum}
\usepackage{fancyhdr}       
\usepackage{graphicx}       
\graphicspath{{media/}}     

\input{header}

\title{Vision Language Model-based Testing of Industrial Autonomous Mobile Robots}

\author{
  Jiahui Wu \\
  Simula Research Laboratory and \\ University of Oslo \\
  Oslo, Norway \\
  \texttt{jiahuiw@uio.no} \\
  \And
  Chengjie Lu \\
  Simula Research Laboratory and \\ University of Oslo \\
  Oslo, Norway \\
  \texttt{chengjielu@simula.no} \\
  \And
  Aitor Arrieta \\
  Mondragon University \\
  Mondragon, Spain \\
  \texttt{aarrieta@mondragon.edu} \\
  \And
  Shaukat Ali \\
  Simula Research Laboratory \\
  Oslo, Norway \\
  \texttt{shaukat@simula.no} \\
  \And
  Thomas Peyrucain \\
  PAL Robotics \\
  Barcelona, Spain \\
  \texttt{thomas.peyrucain@pal-robotics.com} \\
}

\begin{document}
\maketitle

\begin{abstract}
PAL Robotics, in Spain, builds a variety of Autonomous Mobile Robots (AMRs), which are deployed in diverse environments (e.g., warehouses, retail spaces, and offices), where they work alongside humans. Given that human behavior can be unpredictable and that AMRs may not have been trained to handle all possible unknown and uncertain behaviors, it is important to test AMRs under a wide range of human interactions to ensure their safe behavior. Moreover, testing in real environments with actual AMRs and humans is often costly, impractical, and potentially hazardous (e.g., it could result in human injury). To this end, we propose a Vision Language Model (VLM)-based testing approach (\rvsg) for industrial AMRs developed together with PAL Robotics. Based on the functional and safety requirements, \rvsg uses the VLM to generate diverse human behaviors that violate these requirements. We evaluated \rvsg with several requirements and navigation routes in a simulator using the latest AMR from PAL Robotics. 
Our results show that, compared with the baseline, \rvsg can effectively generate requirement-violating scenarios. Moreover, \rvsg-generated scenarios increase variability in robot behavior, thereby helping reveal their uncertain behaviors. 
\end{abstract}

\keywords{Autonomous robots \and Vision language models \and Human behaviors \and Simulation-based testing}

\section{Introduction}
\noindent
PAL Robotics (Spain)~\cite{pal} is a leading service robot manufacturer, with deployments in diverse industries, including healthcare, retail, and agri-food. 
Notably, their Autonomous Mobile Robots (AMRs) excel at safe and efficient navigation in environments, e.g., offices and warehouses, automating tasks alongside humans.
Unpredictable human behavior and frequent interactions between humans and robots place high demands on AMR navigation~\cite{mavrogiannis2023core}. To ensure dependable operation, PAL Robotics defines key functional and safety (F\&S) requirements for various deployments that AMRs must meet.

Scenario-based testing offers a systematic approach to test AMRs' navigation under various human-robot interactions~\cite{ortega2024composable}. Integrating F\&S requirements into automated testing helps generate realistic testing scenarios that violate these requirements~\cite{ortega2024composable}.  
Besides, to improve testing efficiency, a small set of test scenarios that capture diverse human behaviors shall be used, avoiding similar behaviors caused by the same faults~\cite{mullins2018adaptive}.  
Moreover, given the high cost of testing in real environments with actual robots and humans, the impracticality (e.g., asking humans to exhibit a specific behavior), and the potential safety risks, such as collisions that could damage robots or injure humans, simulation-based testing provides a safe and cheap alternative~\cite{afzal2020study}. However, ensuring realistic human behaviors in simulation remains a significant challenge~\cite{afzal2020study}.

To address the challenge mentioned above, we use Vision Language Models (VLMs) to generate test scenarios, which may achieve human-level performance in commonsense reasoning about various tasks~\cite{achiam2023gpt}. 
VLMs can extract key information from both images and text, enabling contextual understanding and drawing logical conclusions. 
Thus, we conjecture that VLMs are naturally well-suited to analyze textual requirements and exhibit strong capabilities in scenario interpretation and construction. This makes VLMs highly capable of analyzing F\&S requirements and generating natural language descriptions of test scenarios and human behaviors. 
An often overlooked but crucial aspect of scenario realism is that human behavior varies depending on the environment. For example, typical human behaviors in warehouses include lifting, sorting packages, and interacting with shelves. 
In simulation-based testing, it is thus essential that modeled human behaviors remain consistent with the logical context of the given environment, which directly influences the configuration of such behaviors. 
VLMs, using their commonsense reasoning capability, can generate contextually appropriate human behaviors tailored to specific environments, which is difficult to achieve with traditional methods, e.g., search algorithms.

To this end, we propose \rvsg, a requirement-driven test scenario generation method for robot navigation tasks based on VLMs having two stages: (i) environment preprocessing and (ii) test scenario generation. 
In the first stage, \rvsg captures and labels images of the simulated environment, then uses the VLM to extract visual features to support scenario generation. 
In the second stage, the prompt generator adapts pre-designed natural language prompt templates based on the given requirement and task. 
Through multi-turn conversations, the VLM generates detailed scenario descriptions and human behavior configurations, which are decoded into executable human agents and run together with the navigation task in the simulator to obtain test results. 
These results are then fed back to the prompt generator to optimize the human configurations, improving test scenarios. 
Moreover, \rvsg uses generated scenarios, human configurations, and feedback to guide the VLM in producing more diverse scenarios.
We evaluated \rvsg and the baseline method on the latest AMR from PAL Robotics using the Gazebo simulator with different requirements and navigation routes. We also employed multiple metrics to assess the performance and diversity of the generated test scenarios.
Our results show that \rvsg can generate more effective requirement-violating test scenarios with a certain degree of diversity, increasing the variability of robot behaviors that help reveal behavioral uncertainties. In addition, different navigation routes affect both the performance of the generated scenarios and the stability of scenario execution.

In summary, our contributions are:
(1) a novel requirement-driven VLM-based test scenario generation method for AMRs;
(2) a set of prompt templates that enable \rvsg to understand the environment, generate test scenarios and human configurations, and dynamically incorporate simulation feedback and historical scenario information for improving scenario generation;
(3) evaluate the test scenarios generated by \rvsg with multiple metrics on the latest AMR from PAL Robotics within the Gazebo simulator, covering a variety of requirements and navigation routes.

\section{Industrial Context}\label{sec:Industrial}
\noindent
This work is conducted in the context of the \EUProject European project~\cite{RoboSAPIENS}, which aims to build and test self-adaptive robots capable of operating safely in uncertain and unknown environments, with adaptation mechanisms implemented as software within the robots. PAL Robotics (Spain)~\cite{pal} is one of the industrial use-case providers in \EUProject. The company is recognized for developing and deploying service robots across various domains, including retail, office, and warehouse environments. Their portfolio includes legged robots, social robots, inventory robots, and Autonomous Mobile Robots (AMRs) for intralogistics applications. 

\begin{figure}[!t]
    \centering
    \resizebox{0.5\linewidth}{!}{\includegraphics{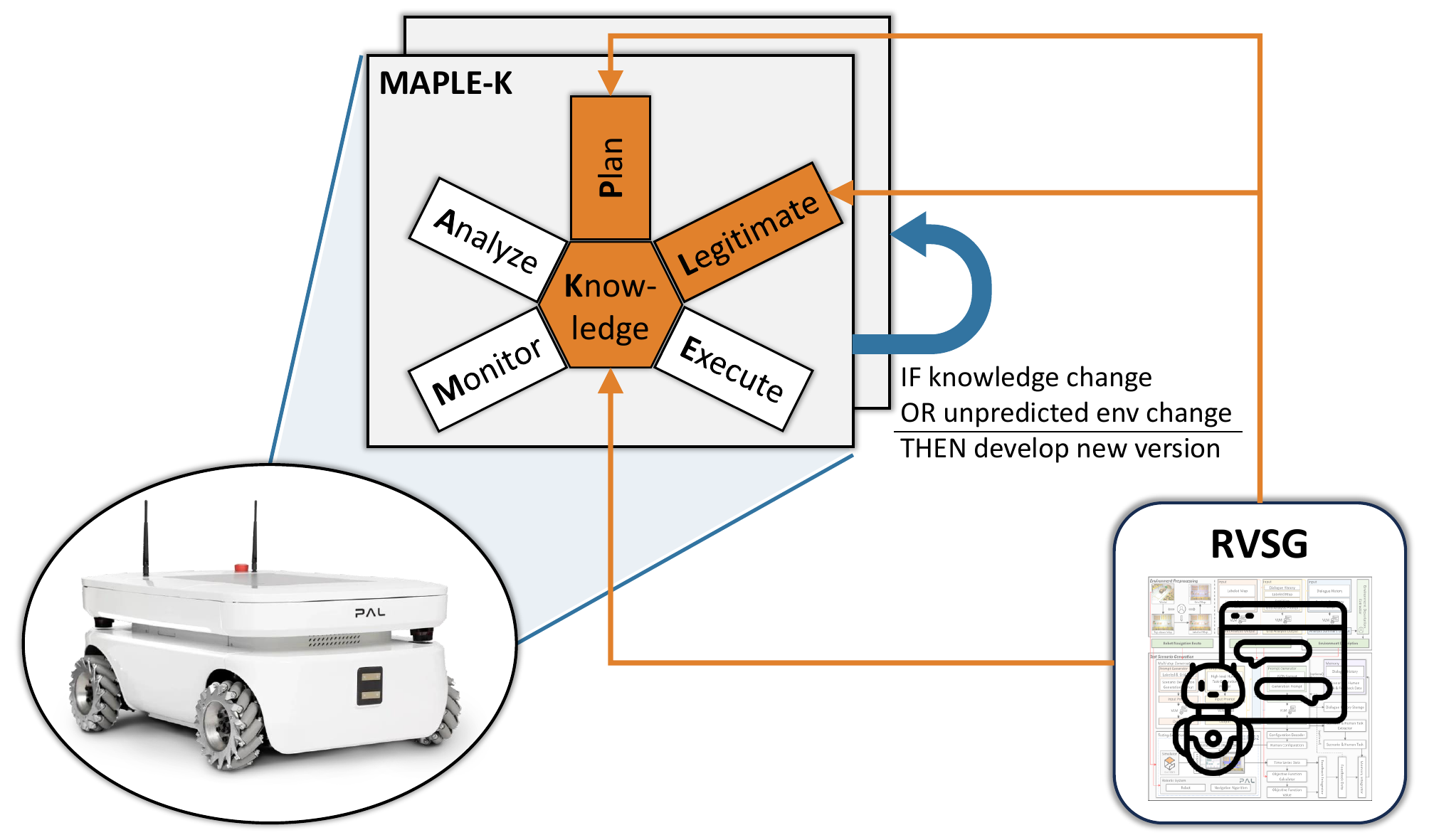}}
    \caption{A representative autonomous mobile robot from PAL Robotics, TIAGo OMNI Base. It is planned to be equipped with the MAPLE-K loop to enable self-adaptation in uncertain and unknown environments. The phases of MAPLE-K highlighted in orange are those supported by \rvsg.}
    \label{fig:maplek}
    \vspace{-15pt}
\end{figure}

This study focuses on AMRs, which are deployed in environments such as offices and warehouses, where they navigate within a defined space and operate alongside humans. Naturally, the interaction between humans and AMRs must be safe. Typically, AMRs use two types of navigation algorithms: a global planner, which determines the optimal path to reach the destination, and a local planner, which accounts for dynamic obstacles in the environment (including humans) and adapts the path accordingly. To this end, an AMR is typically equipped with a set of LiDAR sensors to support various functions such as path planning, collision avoidance, and autonomous navigation. Fig.~\ref{fig:maplek} shows the TIAGo OMNI Base AMR, which is used in many practical applications, both individually and as part of a fleet, to perform its intended operations collaboratively.

Within the \EUProject, self-adaptive behaviors are realized using the MAPLE-K loop~\cite{MAPLE-K}, an extension of the well-established MAPE-K loop for self-adaptation in software systems. As presented in Fig.~\ref{fig:maplek}, in the MAPLE-K loop, an AMR continuously monitors (\textit{M}onitor) its environment using dedicated sensors, analyzes the data (\textit{A}nalyze), and, upon detecting a situation that requires adaptation, plans (\textit{P}lan), validates (\textit{L}egitimate), and executes (\textit{E}xecute) the selected adaptation. Throughout this process, all information becomes part of the robot's knowledge base (\textit{K}nowledge), which can be accessed at any stage of the loop. The \textit{Legitimate} phase is newly introduced in MAPLE-K.

Given the diverse deployment environments of AMRs, it is essential to ensure that they navigate efficiently, stably, and safely while posing no threat to humans~\cite{afzal2020study}. Functional and safety requirements are typically provided by domain experts, based on their domain knowledge and the specifications of both the AMRs and their operating environments. If a potential threat is detected, AMRs adapt their behavior, such as changing their path, stopping, or even communicating with nearby humans. However, due to unpredictable human behavior, it is unknown in advance how humans will interact in the environment where an AMR is deployed. Therefore, it is paramount to test AMRs under a variety of human behaviors, ensuring (to a certain degree) the dependability of their responses. Moreover, conducting tests in real-world environments with real humans is expensive and often impractical~\cite{afzal2020study}. Instead, simulation-based testing offers an affordable alternative, which is the focus of this paper. In general, there is a clear need for systematic methods to test AMR behavior in simulation settings, where diverse human behaviors can be simulated to assess the dependability of AMRs. As Fig.~\ref{fig:maplek} shows, within the context of the MAPLE-K loop, our simulation-based testing approach is designed to support three key phases of MAPLE-K: (1) it enables testing of the MAPLE-K loop implemented in AMRs by generating scenarios that force the AMR to trigger adaptations, thereby ensuring whether the \textit{Plan} component performs its intended functionality; (2) it supports the creation of critical scenarios for the \textit{Legitimate} component, helping assess whether selected adaptations are safe to execute; and (3) it contributes to building a knowledge base of critical scenarios for the \textit{Knowledge} component, contributing to the iterative development of MAPLE-K versions.

\section{Method}\label{sec:Methodology}
\begin{figure}[!htbp]
    \centering
    \resizebox{0.6\linewidth}{!}{\includegraphics{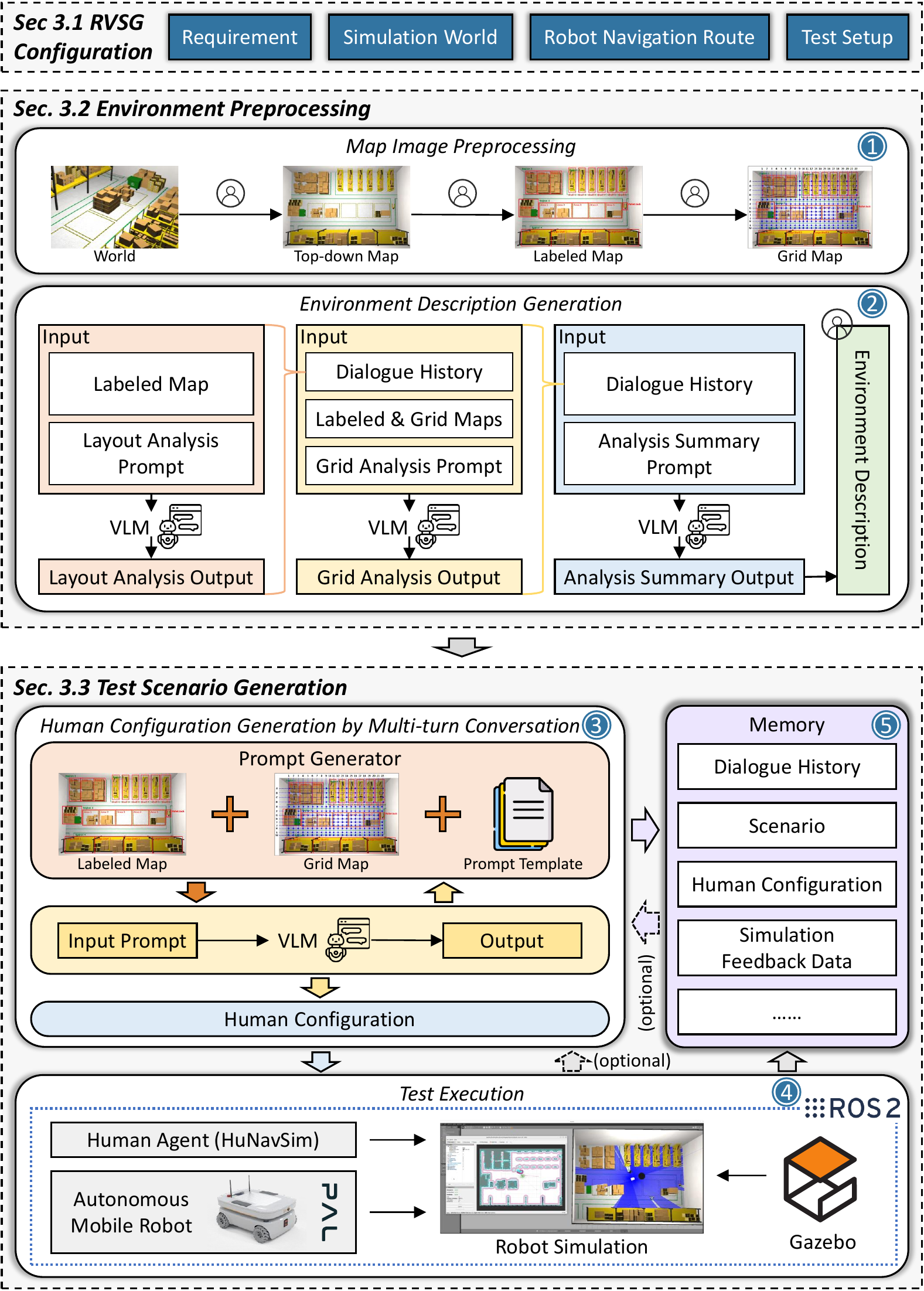}}
    \caption{Overview of \rvsg during the iterative testing of autonomous mobile robots. A process decorated with \humanicon involves a human operator; otherwise, it is fully automated.}
    \label{fig:overview}
    \vspace{-10pt}
\end{figure}

\noindent Fig.~\ref{fig:overview} presents the overview of \rvsg, which aims to generate test scenarios that lead to the robot violating specific requirements. The workflow starts with \textit{\rvsg Configuration} (Section~\ref{subsec:rvsg_config}), where the F\&S requirements, simulation world, and corresponding robot navigation routes are configured. Necessary test setups, such as test budget and stopping criteria, are also defined.
After \rvsg being configured, \textit{Environment Preprocessing} (Section~\ref{subsec:preprocessing}) begins, where \textit{Map Image Preprocessing} is first performed to preprocess the simulation world map (\circlednum{1} in Fig.~\ref{fig:overview}). Based on the preprocessed map, \textit{Environment Description Generation} adopts a VLM to generate natural language descriptions of the environment by a three-turn prompting process at \circlednum{2}. After \textit{Environment Preprocessing} being finished, the preprocessed maps, environment descriptions, together with the requirements and robot navigation routes, and other configurations are forwarded to the next phase for \textit{Test Scenario Generation}.

\textit{Test Scenario Generation} (Section~\ref{subsec:generation}) consists of three key components. The first is \textit{Human Configuration Generation by Multi-turn Conversation} at \circlednum{3}, which generates the human configuration through multi-turn interaction with the VLM. Concretely, this component employs a \textit{Prompt Generator} to construct the input prompts for the VLM. The \textit{Prompt Generator} maintains a set of prompt templates, and can automatically select an appropriate template and instantiate a prompt based on the available information (e.g., preprocessed map images and requirements). 
Once the initial prompt is generated, the multi-turn conversation starts: 
each input–output pair forms a conversation turn and is returned to the \textit{Prompt Generator} as the dialogue history for refining subsequent prompts. The process continues iteratively until all turns are completed, resulting in a final \textit{Human Configuration}. 
Then the \textit{Human Configuration} is simulated in \textit{Test Execution} to test the robot navigation at \circlednum{4}.
The feedback from test execution can be used by \textit{Prompt Generator} to adjust the original prompts for improving effectiveness.
Moreover, we design a \textit{Memory} component to store all multi-turn conversations and test execution feedback collected throughout the experiments (\circlednum{5} in Fig.~\ref{fig:overview}). The \textit{Prompt Generator} leverages this accumulated experience to iteratively refine prompts, increasing diversity in scenario generation.

\subsection{\rvsg Configuration}\label{subsec:rvsg_config}
\noindent
\rvsg Configuration defines the global settings that can be specified in advance for each experiment, including the Requirement, Simulation World, Robot Navigation Route, and Test Setup.
The Requirement (e.g., maintaining a safe distance between the robot and humans) is used to guide \rvsg in generating test scenarios that violate it (e.g., collisions between the robot and humans).
The Simulation World specifies the simulated environment for testing, supporting map acquisition (\circlednum{1} of Fig.~\ref{fig:overview}), simulation execution (\circlednum{4}), and feedback collection (details in Section~\ref{subsubsec:conversation}).
The Robot Navigation Route defines the AMR’s navigation path, including both a natural language description for VLM reasoning and parameter settings for simulation execution (e.g., a set of waypoints for robot navigation).
The Test Setup contains various hyperparameter configurations related to the experiment, with further details provided in Section~\ref{subsec:Setup}.

\subsection{Environment Preprocessing}\label{subsec:preprocessing}

\subsubsection{Map Image Preprocessing}
\rvsg uses a top-down view to capture the image of the simulation world, offering a clear representation of its structural and spatial relationships. 
To further highlight the environment layout, \rvsg labels the image either manually or automatically by drawing bounding boxes around each area and assigning a corresponding name as a label.  
Since the VLM will analyze this labeled image and is highly adaptable in visual analysis, the labeling process is flexible and does not require strict standards, reducing the entry barrier for using \rvsg.
Based on the labeled map image, \rvsg constructs a grid map with precise waypoints, with the adjustable resolution and size tailored to different maps and testing requirements. 
By default, \rvsg uses evenly spaced square grids, where each grid point represents a unique 2D waypoint. For example, if rows are labeled with letters and columns with numbers, a specific waypoint, such as ``E15'', refers to the point in row E and column 15. 
Only valid waypoints (i.e., those not obstructed) are included to aid route planning.
In AMR testing, labeling is typically a one-time effort, allowing the same environment to be reused for testing different robots or for regression testing when their software is updated. As a result, manual labeling is more precise and not necessarily more costly than automated methods (e.g., AI methods), which could produce incorrect or imprecise results. Therefore, \rvsg recommends using manual labeling methods for processing map and grid images.

\subsubsection{Environment Description Generation}
Leveraging the VLM's strong visual reasoning, \rvsg conducts multi-turn conversations to analyze step-by-step the labeled map and grid images, extracting layout and navigation information. 
As shown at \circlednum{2} of Fig.~\ref{fig:overview}, \rvsg first provides the labeled map image and layout analysis instructions to the VLM to obtain a detailed analysis of environmental layout. 
Next, using the preserved dialogue history, \rvsg sends the grid image and corresponding grid analysis instructions to the VLM, along with the map image that emphasizes the basic environmental information,  and generates descriptions of the grid and navigation layouts. 
Finally, \rvsg provides a summarization prompt to the VLM asking to summarize the key description for both map and grid images in JSON format. 
This output is then processed to extract and pair the corresponding descriptions for each image, storing them together as the current environment descriptions for subsequent testing and analysis.
Given the potential for hallucinations or inaccuracies in the VLM's output, the final environment descriptions are manually reviewed to ensure correctness.

\subsection{Test Scenario Generation}\label{subsec:generation}

\subsubsection{Human Configuration Generation by Multi-turn Conversation}\label{subsubsec:conversation}
For \circlednum{3} in Fig.~\ref{fig:overview}, \rvsg utilizes \textit{Prompt Generator} to automatically construct multi-turn conversation prompts based on predefined requirements and robot navigation routes, and preprocessed map images and environment descriptions. 
These prompts are then sequentially input into the VLM, enabling step-by-step reasoning through multi-turn conversations to produce intermediate outputs (e.g., scenario description) and final outputs (e.g., JSON format human configuration). 
Each conversation turn builds on previous context, ensuring the VLM continuously aligns with the test objective, understands the requirement, and generates the required output for each step.
Moreover, by incorporating feedback from scenario simulations (at \circlednum{4}) and memory of previous scenarios (at \circlednum{5}), \rvsg further optimizes configuration generation and enhances scenario diversity. 
Specifically, \rvsg provides multiple prompt templates for five key reasoning steps: (1) \textit{Test Scenario Description Generation}, (2) \textit{High-level Human Task Generation}, (3) \textit{Low-level Human Waypoint Path Generation}, (4) \textit{Human Configuration Validity Check and Revision}, and (5) \textit{JSON Format Human Configuration Generation}.

\begin{itemize}[left=0pt]
\item \textit{Test Scenario Description Generation} analyzes the current environment and produces realistic textual descriptions of test scenarios involving human activities. These descriptions serve as the basis for subsequent human configuration. 
\item \textit{High-level Human Task Generation} allocates specific human tasks according to the generated scenario description, detailing concrete workflows and outputting the names of key labeled areas involved in these workflows. 
\item \textit{Low-level Human Waypoint Path Generation} builds on the high-level tasks, sequentially selecting valid waypoints from the preprocessed grid map based on the workflow and key areas to construct human waypoint paths. 
The high-level tasks and low-level waypoint paths together form the human configuration.
\item \textit{Human Configuration Validity Check and Revision} checks the validity of the generated human configuration. Problematic configurations are automatically detected and revised by the VLM at this stage. For example, the module verifies whether the workflow aligns with the scenario description and whether the waypoint paths are feasible. If inconsistencies are found, the workflow is modified and the waypoint paths are updated accordingly.
\item \textit{JSON Format Human Configuration Generation} provides the final human configuration in JSON format.
\end{itemize}

\begin{figure*}[!t]
    \centering
    \resizebox{\textwidth}{!}{\includegraphics{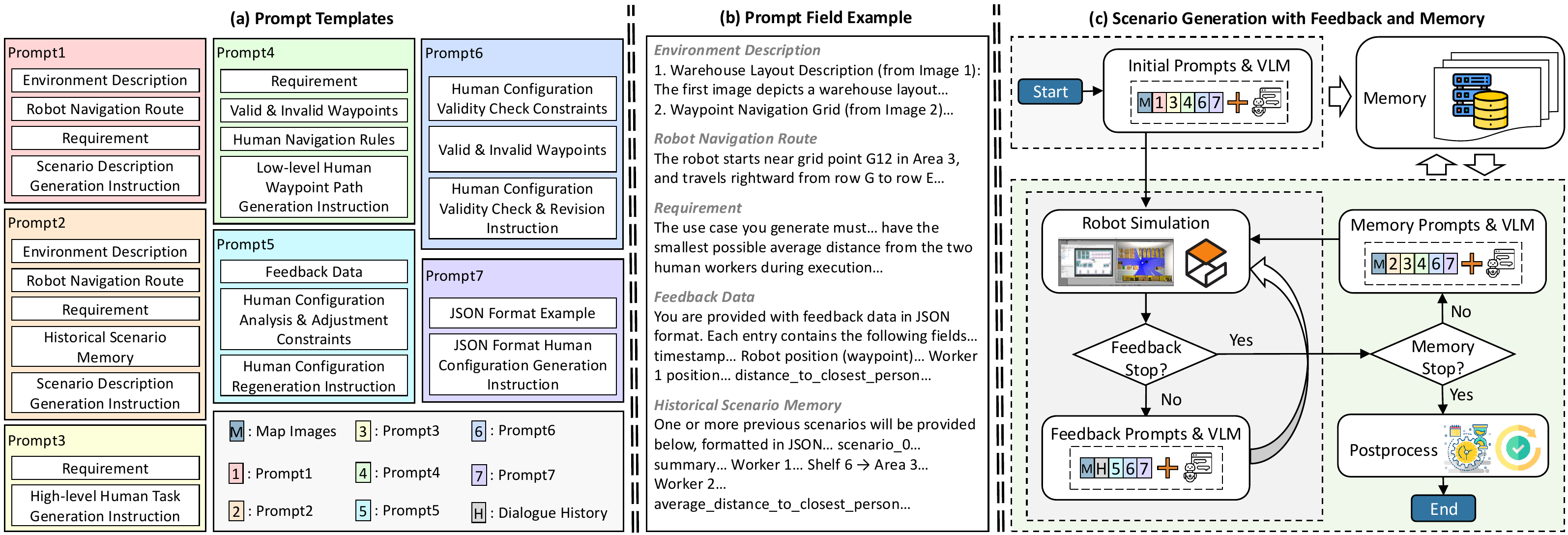}}
    \caption{Prompt templates, prompt field example, and scenario generation with feedback and memory in \rvsg.}
    \label{fig:test_scenario_generation}
    \vspace{-5pt}
\end{figure*}

Fig.~\ref{fig:test_scenario_generation} (a) presents all seven prompt templates provided by \rvsg. Each template focuses on one or two specific key reasoning step(s). 
Based on these templates, \rvsg automatically fills in the provided information, such as requirements, into the corresponding fields to generate prompts for VLM reasoning and analysis. 
As shown in the \textit{Prompt Field Example} in Fig.~\ref{fig:test_scenario_generation} (b), fields such as \textit{Environment Description}, \textit{Robot Navigation Route}, and \textit{Requirement} are filled with content either preprocessed by the VLM or predefined in advance. Fields like \textit{Feedback Data} and \textit{Historical Scenario Memory} are dynamically added or updated during the \rvsg execution, and are filled into the template in JSON format.
Specifically, \textit{Feedback Data} contains time-series information collected at each timestamp during simulation execution, including the positions of the robot and humans, as well as requirement violation metric values (e.g., the minimum distance between the robot and the humans).
\textit{Historical Scenario Memory} includes descriptions of previous scenarios, human configurations, and the corresponding violation metric values obtained in those scenarios.
All templates and instantiated prompts are publicly available in our repository~\cite{RVSG}.

\subsubsection{Test Execution}
At \circlednum{4} in Fig.~\ref{fig:overview}, the \textit{Human Configuration} is used to generate executable \textit{Human Agents} in the simulator. To ensure realistic human behavior, \rvsg employs HuNavSim~\cite{perez2023hunavsim}, an agent controller based on behavior trees and the social force model, which closely simulates real-world human actions. 
The AMR under test is loaded with its designated navigation route and uses its navigation algorithm for environment perception and navigation decisions. 
Both human agents and the AMR perform their tasks in the simulator, enabling dynamic performance evaluation of the AMR along the navigation route, with simulation result data collected and computed throughout the process. 
The simulation results can be used to optimize the prompts at \circlednum{3} for more effective scenario configuration generation. They are also stored in the \textit{Memory} module (at \circlednum{5}) to enable continuous improvement of scenario generation.
The simulation environment is consistent with that used in \textit{Environment Preprocessing}, and real-time communication is maintained via Robot Operating System 2 (ROS2)~\cite{doi:10.1126/scirobotics.abm6074}.

\subsubsection{Memory \& Feedback}
\textit{Memory} module at \circlednum{5} in Fig.~\ref{fig:overview} primarily stores all dialogue history from \circlednum{3}, along with the parsed scenario information and human configurations. It also retains the test execution feedback data from \circlednum{4}. This \textit{Memory} module accumulates experience from scenario generation, enabling iterative optimization of the prompts in \circlednum{3} to support the creation of more effective and diverse scenarios. 
By incorporating test execution feedback from the execution of each generated human configuration in the simulator, \rvsg supports a feedback- and memory-based prompt optimization strategy for the improvement of test scenario generation, as illustrated in Fig.~\ref{fig:test_scenario_generation} (c).

As shown in Fig.~\ref{fig:test_scenario_generation} (c), during the initial generation stage, \rvsg selects a set of prompts, including the initial \textit{Test Scenario Description Generation} prompt (Prompt1), \textit{High-level Human Task Generation} prompt (Prompt3), \textit{Low-level Human Waypoint Path Generation} prompt (Prompt4), \textit{Human Configuration Validity Check and Revision} prompt (Prompt6), and \textit{JSON Format Human Configuration Generation} prompt (Prompt7). These are combined with map images to form the initial multi-turn conversation prompts that guide the VLM in generating the initial human configuration, which is then executed in the simulator to obtain simulation results (e.g., the robot position and the requirement violation metric value at each timestamp). 
Subsequently, the feedback trigger determines whether to initiate feedback-based optimization. If optimization continues, \rvsg combines the dialogue history with map images, \textit{Human Configuration Regeneration} (Prompt5), \textit{Human Configuration Validity Check and Revision} (Prompt6), and \textit{JSON Format Human Configuration Generation} (Prompt7) to guide the VLM in refining the human configuration based on simulation feedback, without altering the initial scenario description. 
The optimized configuration is then executed in the simulator, after which new feedback is obtained. This feedback, including the requirement violation metric value, is used to evaluate the effectiveness of the optimized configuration and determine the degree of optimization. If further optimization is needed, the generated feedback can be used iteratively in subsequent optimization steps. 
If the feedback trigger indicates stopping feedback optimization (e.g., reaching a certain number of optimizations), \rvsg then relies on the memory trigger to decide whether to use memory for generating more diverse scenarios. 
If memory-based optimization proceeds, \rvsg replaces the initial \textit{Test Scenario Description Generation} prompt (Prompt1) with the \textit{Diversity Test Scenario Description Generation} prompt (Prompt2), keeping the other prompts unchanged to guide the VLM in creating the new test scenario. 
Moreover, in Prompt2, the \textit{Historical Scenario Memory} field provides the descriptions of all previously generated test scenarios, along with their corresponding optimal configurations and requirement violation metric values, serving as references for generating more diverse scenarios. 
Otherwise, the process terminates, and all generated scenarios are output.
All multi-turn conversation histories, simulation results, and corresponding scenario descriptions and human configurations are stored in the \textit{Memory} module, supporting VLM-based scenario generation at different stages.

\section{Experiment Design}\label{sec:Design}

\subsection{Subject System and Simulator}\label{subsec:Simulator}
\noindent
We test the latest logistics AMR robot, TIAGo OMNI Base from PAL Robotics~\cite{tiago-omni-base}. 
It has four mecanum wheels for omnidirectional movement and two LiDAR scanners that provide a 360° unobstructed field of view, enabling it to detect and autonomously avoid obstacles to ensure safe navigation. 
To support simulation and testing, PAL Robotics offers an open-source TIAGO OMNI Base ROS 2 Simulation~\cite{palSimulation}. 
The simulator is built on Gazebo classic 11~\cite{gazebo} and provides a high-fidelity model of the TIAGo OMNI Base. 
The TIAGo OMNI Base adopts Nav2~\cite{macenski2020marathon2} for both global navigation and local obstacle avoidance, while real-time communication is achieved through ROS 2 Humble~\cite{doi:10.1126/scirobotics.abm6074}. 
Given the simulation platform and the ROS 2 version, we selected HuNavSim~\cite{perez2023hunavsim} as the default agent for human behavior control. It is capable of reacting to both the robot and humans. We used two human agents in our experiments. 
At the group level, HuNavSim uses the social force model~\cite{helbing1995social} to simulate general crowd flow and avoidance behaviors in the presence of a robot, obstacles, or humans. 
At the individual level, it employs behavior trees to define and control more complex, personalized human actions. 
As a result, HuNavSim can provide realistic human behavior modeling, which benefits AMR navigation task testing and supports the development of navigation systems.

\subsection{Experiment Setup}\label{subsec:Setup}
\noindent
\textit{S\&F Requirements and Objective Functions}. We consider three requirements provided by PAL Robotics: collision avoidance, stability, and efficiency. These requirements reflect essential safety and functional aspects of AMR navigation for PAL robotics and are listed as follows:

\begin{itemize}[left=0pt]
    \item R1: Collision Avoidance – The robot must maintain a safe distance from other objects to prevent collisions.

    \item R2: Stability – The robot must maintain operational stability, ensuring control is not lost due to external forces or loads.

    \item R3: Efficiency – The robot should complete its assigned navigation task within a reasonable time, minimizing delays while maintaining safety and stability.
\end{itemize}

R1 is critical, as the risk of collision is a primary safety concern. To quantify the degree of violation of R1, we opt \textit{Distance to Obstacles} (\dto)~\cite{perez2018teaching} as the objective function:
\begin{equation}
    DTO = \frac{1}{T} \sum_{t=1}^{T} \min_{i=1}^{n} \left\| \mathbf{p}_{\text{robot}}^{t} - \mathbf{p}_{o_i}^{t} \right\|_2,
\end{equation}
where $\mathbf{p}_{\text{robot}}^{t}$ and $\mathbf{p}_{o_i}^{t}$ are positions of the robot and object $o_i$ at timestep $t$, $n$ is the number of objects in the environment, and $T$ is the total number of timesteps.
\dto measures the average minimum Euclidean distance from the robot to the nearest object during its navigation, where smaller values indicate a higher risk of collision.

R2 is also important, as a lack of stability may result in loss of control over the robot. To evaluate the violation of this requirement, we calculate \jerk~\cite{schot1978jerk} as the objective function:
\begin{equation}
    Jerk = \frac{1}{T-1} \sum_{i=1}^{T-1} \frac{\mathbf{a}_{i+1} - \mathbf{a}_i}{\Delta t},
\end{equation}
where $\mathbf{a}_i$ is the acceleration of the robot at timestep $i$, $\Delta t$ is the time interval between steps, and $T$ is the total number of timesteps. \jerk measures the rate of change of acceleration over time; higher jerk values indicate more abrupt motion changes, indicating unstable behaviors.

R3 concerns the time cost required to complete an operational task. To assess the violation of this requirement, we use \textit{Time-to-Reach-Goal} (\trg) as the objective function:
\begin{equation}
    TRG = t_{\text{goal}} - t_{\text{start}},
\end{equation}
where $t_{start}$ and $t_{goal}$ are the timestamps when the robot starts moving and reaches its goal. \trg measures the total duration taken by the robot to finish a given navigation route. Longer \trg indicates inefficiencies or suboptimal behaviors in the robot's execution of the task.

\textit{VLM and Parameter Settings}. We employ GPT-4.1~\cite{openaiGPT-4.1} as the VLM and integrate it into \rvsg for scenario generation. GPT-4.1 is the latest VLM from OpenAI's GPT series, which is designed for complex tasks. It excels at performing a variety of tasks, including coding and instruction following. Moreover, it supports up to 1 million contextual tokens and can better leverage these contexts, improving long-context understanding ability. The parameter settings of VLMs can significantly influence the performance. In our setup, we adopt the default settings from OpenAI for GPT-4.1.

\textit{\rvsg Feedback and Memory Settings}. Based on a pilot study, we set the stopping condition for the feedback trigger to terminate after four configuration optimizations for the same scenario description, resulting in five similar but distinct configurations per scenario. The stopping condition for the memory trigger was set to terminate after generating 10 sets of scenarios, each consisting of one scenario description and its corresponding optimal human configuration. Among these, the scenarios with the best and worst objective value were identified and repeated 30 times for further analysis.

\textit{Comparison Baselines}. Arcuri and Briand suggested~\cite{6032439} to use random search (\textit{RS}) to check if a problem is complex enough to be solved by a complex algorithm. Thus, we performed a pilot study to compare \rvsg with \textit{RS}. Results in our online repository~\cite{RVSG} show that \rvsg is significantly better than \textit{RS} in generating scenarios that violate a given requirement. Hence, our formal experiment excluded \textit{RS} as a baseline. Instead, we employ a modified \rvsg as the comparison baseline, i.e., \rvsgr. \rvsgr shares the same architectural design as \rvsg but excludes the feedback and memory components, resulting in unguided scenario generation by the VLM.

\begin{figure}[!t]
    \centering
    \resizebox{0.5\linewidth}{!}{\includegraphics{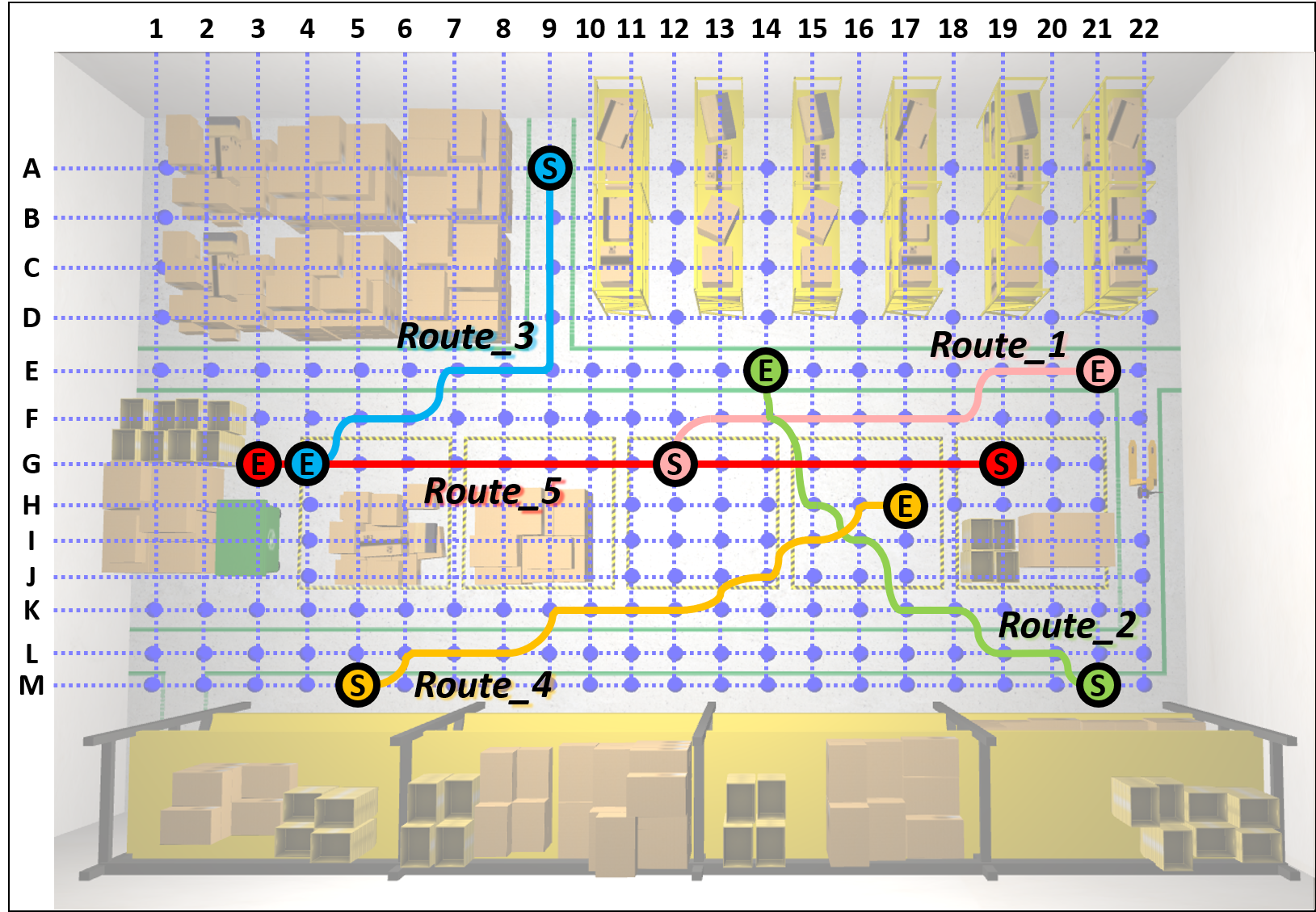}}
    \caption{Warehouse simulation world and navigation routes for AMR testing. Circles labeled with ``S" and  ``E" indicate the starting and end of the routes, respectively.}
    \label{fig:route}
    \vspace{-5pt}
\end{figure}

\textit{Navigation Routes}. We used the Amazon warehouse simulation world~\cite{awsWarehouse} as the test environment for our experiments. As shown in Fig.~\ref{fig:route}, the environment consists of multiple functional areas, with shelves, boxes, and other obstacles distributed across different regions, providing a complex and realistic simulation for AMR testing.
We selected five different navigation routes that cover the entire warehouse. Some routes represent safer navigation scenarios. For example, in \textit{Route\_1}, the robot travels primarily through open areas with almost no nearby obstacles. Other routes include more challenging situations. For instance, in \textit{Route\_3}, the robot navigates very close to boxes and other obstacles in certain segments. This diverse set of routes enables a more comprehensive evaluation of the effectiveness of the generated test scenarios.

\subsection{Evaluation Metrics}\label{subsec:Metrics}
\noindent
\textit{Objective Metric}. We adopted the same three objective functions as in Section~\ref{subsec:Setup}, i.e., \dto, \jerk, and \trg, as the objective evaluation metrics to directly assess the effectiveness of the generated test scenarios in violating different requirements. 

\textit{Performance Metric}. We incorporated four metrics from~\cite{perez2023hunavsim} to provide a performance assessment of the scenarios generated by \rvsg for various requirement violations:

\begin{itemize}[left=0pt]
\item \textit{Robot on Person Collision} (\rpc): count the number of times the robot collides with a person. This metric is calculated using the positions, orientations, and velocities of both the robot and humans to determine collision events and identify which party initiates the collision.
\item \textit{Cumulative Heading Changes} (\chc): measures the cumulative heading changes along the robot’s navigation, calculated as the sum of angles between successive waypoints. It reflects both path smoothness and motion stability, with higher values indicating less smooth and less stable movement.
\item \textit{Path Length} (\pl): measures the distance the robot travels from the initial point to the goal position, which is used to assess navigation efficiency, with longer paths generally indicating less efficient planning.
\item \textit{Time Not Moving} (\tnm): quantifies the seconds the robot remains stationary, which is used to evaluate the efficiency and continuity of the robot’s movement, with higher values indicating less continuous and effective navigation.
\end{itemize}

\textit{Diversity Metric}. To evaluate the diversity of the scenarios generated by \rvsg, we defined the following metrics:

\begin{itemize}[left=0pt]
\item \textit{Scenario Description Diversity} (\sdd): measures the diversity of natural language scenario descriptions, defined as one minus the average similarity computed by Sentence Transformers~\cite{reimers-2019-sentence-bert}. Higher values indicate greater diversity in scenario descriptions.
\item \textit{High-level Task Diversity} (\htd): measures the diversity of human high-level task descriptions by computing the average Levenshtein distance~\cite{lcvenshtcin1966binary} between them, which quantifies how many single-character edits (insertions, deletions, or substitutions) are needed to change one string into another. Higher values indicate greater diversity.
\item \textit{Scenario Simulation Diversity} (\ssd): measures the diversity of simulated scenarios by computing the average dynamic time warping (DTW) distance~\cite{berndt1994using} between the time-series position data of the robot and humans. DTW calculates the distance between two temporal sequences, with larger distances indicating greater dissimilarity. Thus, higher values reflect greater diversity among simulated scenarios.
\end{itemize}

\subsection{Research Questions}\label{subsec:RQs}
\noindent
We addressed the following research questions (RQs): \textbf{RQ1}: How does \rvsg perform compared to \rvsgr on different navigation routes? \textbf{RQ2}: How do \rvsg-generated scenarios differ in exposing variability and instability in robot behavior? \textbf{RQ3}: How do navigation routes affect the performance and variability of \rvsg-generated scenarios?

\subsection{Statistical Tests}\label{subsec:Statistical}
\noindent
Since our data are continuous and may not satisfy the normality assumption, we chose the following guide~\cite{arcuri2011practical} to select the \textit{Mann-Whitney U test} for determining statistical significance of our results (significance level 0.05). Effect sizes were assessed with the \textit{Vargha and Delaney \Atwelve} statistic, with values of $\geq$ 0.638 or $\leq$ 0.362 considered indicative of medium or large effect sizes based on the guides~\cite{arcuri2011practical,kitchenham2017robust}.

\section{Results and Analysis}\label{sec:Results}

\subsection{Comparison between \rvsg and \rvsgr --- RQ1}\label{subsec:RQ1_results}

\begin{table*}[!htbp]
    \centering
    \caption{Results achieved by \rvsg and \rvsgr for all metrics on different routes and requirements. ``A/B" means the results from \rvsgr and \rvsg. A is from \rvsgr and B is from \rvsg. \textbf{Bold} indicates a method with better performance. -- \textbf{RQ1}}
    \resizebox{\linewidth}{!}{\input{tables/tabRQ1}}
    \label{tab:tabRQ1}
\end{table*}

\noindent 
Table~\ref{tab:tabRQ1} shows the results of \rvsg and \rvsgr across all metrics and navigation routes. We denote \rvsg under three requirement settings as \rvsgd, \rvsgj, and \rvsgt. When looking at \textit{Objective Metric}, for \textit{R1}, \rvsgd consistently outperformed \rvsgr across all routes in \dto, indicating more effective violation of \textit{R1} by \rvsg than \rvsgr. For \textit{R2}, \rvsgj led to higher jerk on four routes (i.e., \textit{\rone-4}), showing better violation of \textit{R2} by \rvsg than \rvsgr. For \textit{R3}, \rvsgt outperformed \rvsgr on all routes, meaning \rvsgt generated scenarios that effectively hindered the robot from reaching its goal.

For \textit{Performance Metric}, \rvsgd outperformed \rvsgr in \rpc, showing that violating \textit{R1} increases the chances of robot-responsible collisions. However, \rvsgj and \rvsgt, led to higher \rpc in only about half of the cases, since \textit{R2} and \textit{R3} are not related to collision and thus provide weaker guidance for collision than \rvsgd. For \chc and \pl, \rvsg outperformed \rvsgr in 27 out of 30 cases across all requirements and routes. In the remaining three cases, differences were marginal (i.e., 1.386 vs. 1.311, 8.482 vs. 8.472, 8.482 vs. 8.477), indicating that \rvsg consistently generates scenarios that effectively degrade navigation performance. For \tnm, the performance between \rvsg and \rvsgr was comparable; in certain cases, \rvsg performed better, while in others, it performed slightly worse. Scenario replay showed that higher \tnm does not always reflect worse performance, as it can also represent appropriate pauses or safe waiting behaviors.

For \textit{Diversity Metric}, for \sdd and \htd, \rvsg outperformed \rvsgr across all requirements and routes, showing greater diversity in scenario descriptions and high-level task descriptions. However, for \ssd, \rvsg outperformed \rvsgr in 7 of 15 cases (0 for \textit{R1}, 3 for \textit{R2}, 4 for \textit{R3}). This suggests that \rvsg's strength lies in linguistic and task-level diversity rather than in producing diverse human-robot interactions. A possible reason is that \rvsgr, resembling random search, explores the search space more broadly.

\begin{center}
    \fcolorbox{black}{gray!10}{
    \parbox{0.96\columnwidth}{
    \textbf{Conclusion for RQ1}: 
    \rvsg outperformed \rvsgr in generating scenarios that violate safety and performance requirements, degrading robot performance, and enhancing linguistic and task-level diversity. However, \rvsg's ability to simulate diverse scenarios is limited, likely due to its more guided search strategy.
    }}
\end{center}

\subsection{Assessing Robot Behavioral Variability and Instability with \rvsg-Generated Scenarios --- RQ2}\label{subsec:RQ2_results}

\begin{table*}[t]
    \centering
    \caption{Statistical test, standard deviation, and diversity results for the best and worst scenarios generated by \rvsg (repeated 30 times) across different metrics and requirements. ``/" denotes no statistical results. \textbf{Bold} indicates a significant result. -- \textbf{RQ2}}
    \resizebox{0.9\linewidth}{!}{\input{tables/tabRQ2}}
    \label{tab:tabRQ2}
    \vspace{-5pt}
\end{table*}

\noindent
We picked the best and worst scenarios from all the scenarios generated by \rvsg. We executed each selected scenario 30 times to study its characteristics for variability in the robot's behaviors. For \textit{Objective Metric} and \textit{Performance Metric}, we performed the \textit{Mann-Whitney U test} and the \textit{Vargha and Delaney \Atwelve} effect size to compare differences between all the best (\rvsgb) and worst (\rvsgw) scenarios. We also calculated the standard deviation (\textit{std}) for these metrics to understand the variability in the robot's behavior across repeated executions. 
For \textit{Diversity Metric}, we calculated \textit{\ssd} across the 30 simulations for each scenario. Table~\ref{tab:tabRQ2} summarizes the results for different requirements. For \textit{R1}, results showed no statistically significant difference between \rvsgb and \rvsgw in all metrics except for \dto, indicating similar robot performance. Nevertheless, for \dto, \rvsgb significantly outperformed \rvsgw, meaning it better violated R1. Moreover, \rvsgb consistently achieved higher \textit{std} values for almost all metrics. This indicates less stable robot behaviors when using \rvsgb compared to \rvsgw, suggesting that \rvsgb is more suitable for uncovering unexpected behaviors during testing. For \textit{R2}, \rvsgb achieved a comparable performance as \rvsgw for five out of seven metrics (i.e., \jerk, \trg, \rpc, \pl, \tnm), and for the other two metrics (\dto, \chc), \rvsgb significantly outperformed \rvsgw. Regarding variability, \rvsgb consistently achieved higher \textit{std} values across all metrics, indicating less stable robot behaviors during testing. Similarly, for \textit{R3}, \rvsgb achieved comparable or significantly better performance than \rvsgw across the metrics, while also leading the robot to exhibit more unstable behaviors. 
This consistent pattern indicates the best scenarios identified by \rvsg are more effective at exposing robot's unexpected or unstable behaviors, making them more suitable for testing. Finally, the diversity results (i.e., \ssd) show that \rvsgb performed more unstably than \rvsgw for \textit{R1} and \textit{R3}, while for \textit{R2}, \rvsgb is more stable than \rvsgw. This variation may lie in the requirements, and further experiments are needed to confirm this hypothesis.

\begin{center}
    \fcolorbox{black}{gray!10}{
    \parbox{0.96\columnwidth}{
    \textbf{Conclusion for RQ2}: 
    \rvsgb induces more unstable robot behaviors than \rvsgw across different requirements. Besides, \rvsgb is consistently more effective than \rvsgw and, in several cases, significantly outperforms \rvsgw on certain metrics such as \dto and \chc. 
    }}
\end{center}

\subsection{Studying the Effect of Navigation Routes in \rvsg-Generated Scenarios --- RQ3}\label{subsec:RQ3_results}

\begin{figure}[t]
    \centering
    \resizebox{0.6\linewidth}{!}{\includegraphics{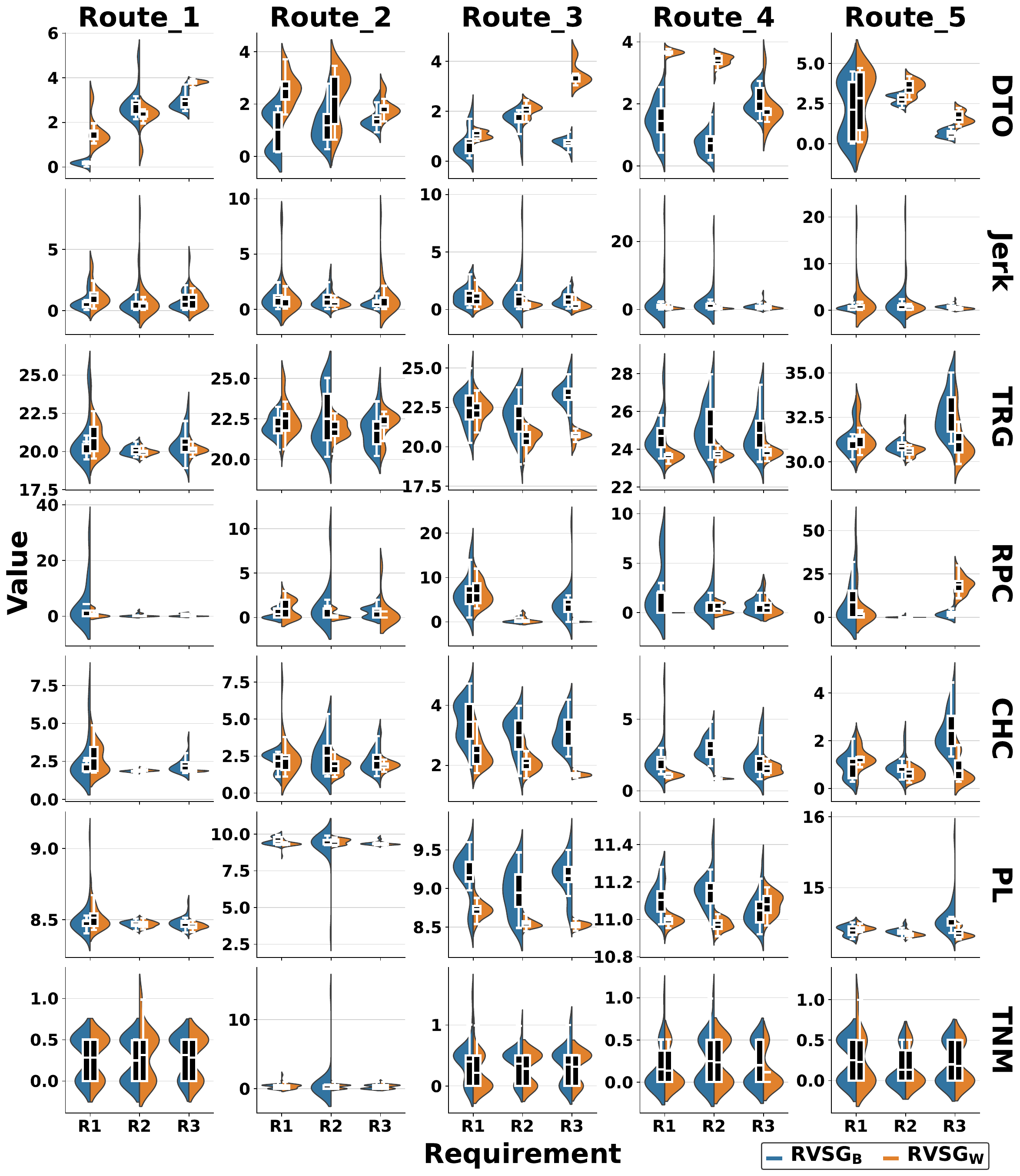}}
    \caption{Distribution results of the best and worst scenarios (repeated 30 times) generated by \rvsg with different requirements for different metrics under different navigation routes. The mean represents the central tendency. -- \textbf{RQ3}}
    \label{fig:plot}
    \vspace{-10pt}
\end{figure}

\noindent
We further investigated the impact of different routes on the performance of the test scenarios generated by \rvsg with the same settings as RQ2 (i.e., picking the best and worst scenarios and the same statistics). 
We summarize the key findings here, and all detailed results are available in our public repository~\cite{RVSG}.

For R1, \rvsgb significantly outperformed \rvsgw for \dto on four out of the five routes, with the only exception being \rfive. On \rfive, no significant differences were observed across all metrics. Notably, \rfive is a straight path without any turns, resulting in a lower task complexity. This may diminish the performance differences between the two scenarios, leading to the absence of significant differences in the results.
For other metrics, \rvsgb was only significantly worse than \rvsgw in a few cases. However, changes in the navigation route could have also affected the results. For example, in \rone, \rvsgb performed significantly worse than \rvsgw for \jerk, while in \rfour, it performed better. This may be because the generated test scenarios, combined with routes passing close to shelves and boxes (as in \rfour), introduce additional complexities. These factors further restrict the robot's movement space and require more frequent adjustments, thereby impacting the \jerk values.
In addition, we observe that in \rthree and \rfour, \rvsgb had a higher \textit{std} than \rvsgw. For other routes, the \textit{std} for different metrics varied. This suggests that the navigation route affects the stability of scenario execution and leads to different levels of variability.
For R2, \rvsgb performed significantly worse than \rvsgw in only two metrics (\dto and \chc) in \rone. However, in other routes, \rvsgb significantly outperformed \rvsgw in terms of these two metrics. Regarding the \textit{std}, \rvsgb showed higher values than \rvsgw in some routes and lower values in others. This suggests that the navigation route influences the performance of scenarios generated by \rvsg.
For R3, \rvsgb outperformed \rvsgw or showed no significant difference in most metrics. The \textit{std} of \rvsgb was higher than \rvsgw in some routes and lower in others, indicating that the route affects the performance of \rvsg. 
For \ssd, \rvsgb exhibited greater diversity than \rvsgw in some routes but less in others, suggesting influence by the route.
Fig.~\ref{fig:plot} shows the distribution results of \rvsgb and \rvsgw for different metrics and requirements across various routes. Overall, the navigation routes influenced the distribution of both scenarios for the same requirement or metric. Moreover, the figure indicates that in \rthree and \rfour, \rvsgb exhibited higher distribution than \rvsgw, suggesting that the robot displays greater behavioral variability in scenarios produced by \rvsgb.

\begin{center}
    \fcolorbox{black}{gray!10}{
    \parbox{0.96\columnwidth}{
    \textbf{Conclusion for RQ3}: 
    Navigation routes affect the performance of \rvsgb and \rvsgw under different requirements and metrics, as well as the stability of scenario execution and the robot's behavioral diversity. Different routes may lead to significant changes in metrics' values and distributions, highlighting the crucial role of navigation routes.
    }}
\end{center}

\section{Threats to Validity}\label{sec:Threats}
\noindent
Concerning \textit{conclusion validity concerns}, to draw a reliable conclusion, we employed appropriate statistical tests and a rigorous analysis based on established guides. 
Regarding \textit{internal validity} concerns related to parameters, we used the default parameters for the VLM from OpenAI and kept the same parameter settings for all approaches.
We also repeated each scenario 30 times to mitigate randomness in the simulator and ensure stable results.
Concerning \textit{constructive validity} threat, we used human-aware navigation metrics from HuNavSim~\cite{perez2023hunavsim}. These metrics are comparable and commonly applied in robot navigation~\cite{tsoi2022sean,biswas2022socnavbench}. To mitigate threats due to hallucinations and errors in VLM outputs, we manually reviewed a subset of the generated responses.

An \textit{External Validity} threat in our evaluation relates to the generalizability of the results. The first threat concerns obtaining the same level of benefits observed in our evaluation, with additional requirements, environments, and routes using TIAGO OMNI Base. To this end, we used three requirements, five robot navigation routes, and one environment, all of which were selected in conjunction with PAL to reflect real-world operations and human-robot interactions. Nonetheless, more experiments are warranted in the future. The second is the applicability of \rvsg to other AMRs. To this end, experiments with additional AMRs are needed. However, note that we used the industrial AMR in a real industrial setting provided by PAL. Third, to address the simulator fidelity threat, we used Gazebo, a high-fidelity simulator that enables realistic modeling of robot dynamics, sensor noise, and human-agent interactions. Moreover, this simulator is used by PAL in their practice. Besides, to enhance the realism of human behaviors, we used HuNavSim~\cite{perez2023hunavsim}, which integrates the social force model and behavior trees, and diverse human reactions to robot presence.

\section{Lessons Learned and Industrial Perspectives}\label{sec:Lessons}

\noindent

\noindent\textbf{Lesson 1 -- Requirements-driven guidance improves scenario relevance and diversity:} Compared to unguided generation, our requirement-driven VLM-based test generation method consistently led to scenarios that not only violated the intended system requirements but also exposed diverse instabilities in our industrial AMR behavior in a realistic manner. This highlights the importance of aligning VLMs with domain-specific requirements rather than relying solely on open-ended prompting.

\noindent\textbf{Lesson 2 -- VLM-generated human behaviors increase simulation realism and fidelity: }While Gazebo and HuNavSim~\cite{perez2023hunavsim} provide accurate physical and behavioral simulations, our technique makes the scenarios realistic and challenging in terms of human behaviors, thanks to the integration of VLM-based methods. This underscores that when testing AMRs, achieving fidelity is not only about physical fidelity but also about contextual realism of simulated entities.

\noindent\textbf{Lesson 3 -- Feedback and memory mechanisms enhance test diversity:} Integrating simulation feedback and historical memory into \rvsg improved linguistic and task-level diversity, producing better scenarios over time. However, this guided search strategy reduced the coverage of rare but potentially critical edge cases, which unguided approaches sometimes revealed. A hybrid strategy that combines guided and exploratory generation is envisioned as a future research avenue.

\noindent\textbf{Lesson 4 -- Navigation routes critically shape test outcomes: }Our evaluation showed that scenario effectiveness varies depending on different navigation routes. While straight routes yielded fewer violations, routes with environmental complexities (e.g., routes close to shelves) increased requirement violations and instability. Our discussion with practitioners reinforced this finding, as they emphasized that realistic routes often include such complex and constrained layouts, making route diversity a critical factor when generating test cases.

\noindent\textbf{Lesson 5 -- \rvsg supports the MAPLE-K loop enabling self-adaptation in robots:} PAL Robotics is implementing the MAPLE-K loop inside its AMRs to allow them to deal with unsafe situations with self-adaptation. The work presented in this paper was carried out in collaboration with engineers from PAL, resulting in a tool that they will use to support various phases of MAPLE-K, including Plan, Legitimate, and Knowledge. PAL will utilize this tool for AMRs as well as other types of robots they are developing, and consider this tool an important addition to their software engineering tool suite for self-adaptive robotics.

\noindent\textbf{Lesson 6 -- Industrial adoption requires balancing automation with expert oversight: }Discussions with our industrial partners confirmed that, while RVSG substantially reduces manual effort, lightweight expert validation is still recommended to ensure no hallucination, confirming, this way, scenario feasibility. While VLM-based testing can scale scenario generation, deployment at scale may still require a human-in-the-loop role to ensure both trustworthiness and acceptance in practice.

\section{Related Works}\label{sec:Related}

\noindent
\textit{Testing of Robotic Systems}.
Various robotic testing approaches have been proposed, e.g., simulation-based testing~\cite{9438553,timperley2018crashing,lu2024epitester,santos2018property} and formal verification~\cite{luckcuck2019formal,miyazawa2019robochart}. For example, Hutchison et al.~\cite{hutchison2018robustness} proposed a framework to automate the test generation and execution for industrial autonomy systems, including robots, and robotics-oriented libraries. 
Brito et al.~\cite{brito2022integration} proposed an integration testing approach for robotic systems focusing on the publish/subscribe communication function and generating critical test scenarios guided by coverage criteria. 
To ensure the correct ROS operation, Santos et al.~\cite{santos2018property} presented a property-based testing approach for ROS system configurations, focusing on orchestration errors that occur when integrating different ROS components at runtime. 
To support the modeling and verification of robotic applications, Miyazawa et al.~\cite{miyazawa2019robochart} proposed RoboChart, a domain-specific modeling language based on UML. 
RoboChart supports modeling and verification of robotics applications through model checking and theorem proving.
Compared to the above methods, \rvsg focuses on system-level testing of an industrial robot to generate scenarios including human behaviors in simulation. It uses a VLM to search for semantically meaningful and realistic scenarios that challenge the robot's behavior to violate its functional and safety requirements.

\noindent 
\textit{Foundation Models-based Test Generation}. 
Foundation models (FMs), e.g., LLMs and VLMs, have shown benefits in many areas. Among them, FM-based test generation offers an effective way to generate tests~\cite {wang2024software,hou2024large,10329992}. With techniques, e.g., prompting or fine-tuning, FMs support testing activities, including unit testing~\cite{10329992,tufano2020unit} and test oracle generation~\cite{tufano2022generating,nie2023learning}. 
For instance, Schäfer et al.~\cite{10329992} proposed TESTPILOT, an adaptive LLM-based test generation tool for JavaScript. 
TESTPILOT generates unit tests by iteratively querying an LLM with a specifically designed prompt. 
FMs also show promise in test minimization. For example, Pan et al.~\cite{pan2024ltm} proposed LTM, a test suite minimization approach based on LLMs and similarity measures.
Though FMs showed good performance in testing tasks like unit testing, their use to test complex software systems, e.g., robots, remains unexplored due to their complexity (e.g., real-time constraints and multimodal inputs), which pose unique testing challenges. Some works address these challenges for autonomous vehicles with FMs~\cite{lu2024multimodal,lu2024realistic}. 
In contrast, \rvsg generates test scenarios for an industrial robot operating in dynamic environments.
Specifically, \rvsg focuses on scenarios involving human-robot interactions, which are critical to both functionality and safety. Using VLMs, \rvsg can synthesize realistic scenarios that expose potential violations of robots' safety and functional requirements.


\section{Conclusions and Future Work}\label{sec:Conclusions}
\noindent
We present \rvsg, a VLM-based test scenario generation approach for robot navigation. Given a functional or safety requirement, \rvsg generates scenarios violating the requirement. \rvsg employs a VLM to extract visual features from an image describing a robot's environment. Next, the extracted features are then used by the VLM to generate human behavior configurations along with the scenario description by iteratively querying the VLM with specifically designed prompts. The scenarios generated in each iteration are then executed in simulation, and the execution results are used as feedback for the next iteration to refine the scenarios. We evaluated \rvsg in a warehouse environment and selected three requirements and five robot navigation routes for the experiment. The results show that \rvsg can generate effective requirement-violating test scenarios that increase the variability of robot behaviors, helping reveal behavioral uncertainties. Meanwhile, the navigation routes affect the performance and stability of \rvsg-generated scenarios. 
Future works include studying the generalizability of \rvsg by conducting experiments with other robots. Also, we plan to investigate multi-objective \rvsg that can violate multiple requirements simultaneously.
Our repository is now publicly available online~\cite{RVSG}.

\section*{Acknowledgments}
This work is supported by the RoboSAPIENS project funded by the European Commission’s Horizon Europe programme (grant No. 101133807), and by the Co-tester (No. 314544) and Co-evolver (No. 286898/F20) projects funded by the Research Council of Norway. Aitor Arrieta is part of the Software and Systems Engineering research group of Mondragon Unibertsitatea (IT1519-22), supported by the Department of Education, Universities and Research of the Basque Country. Aitor Arrieta is also supported by the Spanish Ministry of Science, Innovation and Universities (project PID2023-152979OA-I00), funded by MCIU /AEI /10.13039/501100011033 / FEDER, UE.

\bibliographystyle{unsrt}  
\bibliography{references}

\end{document}

%% file: header.tex
\usepackage{comment}
\usepackage{balance}
\usepackage{url}
\usepackage{subcaption}
\usepackage{booktabs}
\usepackage{multirow}
\usepackage{xspace}
\usepackage{enumitem}
\usepackage{bm}
\usepackage{tikz}
\usepackage{amsmath}

\newcommand{\circlednum}[1]{%
  \tikz[baseline=(char.base)]{
    \node[shape=circle,draw,inner sep=0.5pt, line width=0.3pt, scale=1] (char) {\sffamily\bfseries\footnotesize #1};
  }%
}

\newcommand{\pl}{\textit{PL}\xspace}
\newcommand{\chc}{\textit{CHC}\xspace}
\newcommand{\rpc}{\textit{RPC}\xspace}
\newcommand{\tnm}{\textit{TNM}\xspace}

\newcommand{\humanicon}{\includegraphics[height=2.6ex]{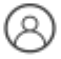}\xspace}

\newcommand{\dto}{\textit{DTO}\xspace}
\newcommand{\jerk}{\textit{Jerk}\xspace}
\newcommand{\trg}{\textit{TRG}\xspace}

\newcommand{\sdd}{\textit{SDD}\xspace}
\newcommand{\htd}{\textit{HTD}\xspace}
\newcommand{\ssd}{\textit{SSD}\xspace}

\newcommand{\rone}{\textit{Route\_1}\xspace}
\newcommand{\rtwo}{\textit{Route\_2}\xspace}
\newcommand{\rthree}{\textit{Route\_3}\xspace}
\newcommand{\rfour}{\textit{Route\_4}\xspace}
\newcommand{\rfive}{\textit{Route\_5}\xspace}

\newcommand{\rvsg}{\textit{RVSG}\xspace}
\newcommand{\rvsgr}{$\rvsg_{R}$\xspace}
\newcommand{\rvsgd}{$\rvsg_{\dto}$\xspace}
\newcommand{\rvsgj}{$\rvsg_{\jerk}$\xspace}
\newcommand{\rvsgt}{$\rvsg_{\trg}$\xspace}
\newcommand{\rvsgb}{$\rvsg_{B}$\xspace}
\newcommand{\rvsgw}{$\rvsg_{W}$\xspace}

\newcommand{\Atwelve}{\^{A}\textsubscript{12}\xspace}

\newcommand{\EUProject}{{\tt RoboSAPIENS}\xspace}

%% file: tables/tabRQ1.tex
\begin{tabular}{llllllllll}
\toprule
\multirow{2.5}{*}{\begin{tabular}[c]{@{}l@{}}\textbf{Requirement}\\ \textbf{(Objective)}\end{tabular}} & \multirow{2.5}{*}{\textbf{Route}} & \multirow{2.5}{*}{\textbf{Objective Metric}} & \multicolumn{4}{l}{\textbf{Performance Metric}} & \multicolumn{3}{l}{\textbf{Diversity Metric}} \\ \cmidrule(r){4-7} \cmidrule(r){8-10} 
 &  &  & \textbf{\rpc$\uparrow$} & \textbf{\chc$\uparrow$} & \textbf{\pl$\uparrow$} & \textbf{\tnm$\uparrow$} & \textbf{\sdd$\uparrow$} & \textbf{\htd$\uparrow$} & \textbf{\ssd$\uparrow$} \\ \midrule
\multirow{5}{*}{\begin{tabular}[c]{@{}l@{}}Collision \\Avoidance\\ (\dto$\downarrow$)\end{tabular}} & \rone & 3.020 / \textbf{0.666} & 0.2 / \textbf{3.7} & 2.002 / \textbf{2.222} & 8.482 / \textbf{8.491} & 0.249 / \textbf{0.299} & 0.142 / \textbf{0.253} & 4.550 / \textbf{4.950} & \textbf{60.415} / 41.385 \\
 & \rtwo & 3.717 / \textbf{0.809} & 0.1 / \textbf{1.9} & 1.345 / \textbf{2.072} & 9.279 / \textbf{9.398} & 0.299 / \textbf{0.351} & 0.127 / \textbf{0.209} & 4.744 / \textbf{6.244} & \textbf{70.531} / 39.966 \\
 & \rthree & 1.950 / \textbf{0.390} & 0.7 / \textbf{8.0} & 2.275 / \textbf{2.837} & 8.136 / \textbf{8.934} & \textbf{1.849} / 0.250 & 0.136 / \textbf{0.234} & 4.611 / \textbf{4.783} & \textbf{55.667} / 27.019 \\
 & \rfour & 3.433 / \textbf{1.432} & 1.2 / \textbf{1.3} & 1.036 / \textbf{2.034} & 10.996 / \textbf{11.118} & 0.301 / \textbf{0.351} & 0.148 / \textbf{0.187} & 4.067 / \textbf{6.228} & \textbf{56.523} / 51.388 \\
 & \rfive & 3.226 / \textbf{1.501} & 3.5 / \textbf{5.3} & 1.386 / \textbf{1.969} & 14.411 / \textbf{14.656} & \textbf{0.201} / 0.200 & 0.127 / \textbf{0.228} & 4.217 / \textbf{5.928} & \textbf{69.550} / 51.151 \\ \midrule
\multirow{5}{*}{\begin{tabular}[c]{@{}l@{}}Stability\\ (\jerk$\uparrow$)\end{tabular}} & \rone & 0.980 / \textbf{1.371} & 0.2 / \textbf{0.3} & 2.002 / \textbf{2.020} & \textbf{8.482} / 8.472 & \textbf{0.249} / 0.248 & 0.142 / \textbf{0.200} & 4.550 / \textbf{8.228} & 60.415 / \textbf{61.838} \\
 & \rtwo & 0.613 / \textbf{1.354} & 0.1 / \textbf{1.0} & 1.345 / \textbf{1.813} & 9.279 / \textbf{9.366} & 0.299 / \textbf{0.449} & 0.127 / \textbf{0.200} & 4.744 / \textbf{8.522} & \textbf{70.531} / 61.546 \\
 & \rthree & 0.949 / \textbf{2.239} & 0.7 / \textbf{1.8} & 2.275 / \textbf{3.051} & 8.136 / \textbf{8.941} & \textbf{1.849} / 0.399 & 0.136 / \textbf{0.188} & 4.611 / \textbf{8.706} & \textbf{55.667} / 45.093 \\
 & \rfour & 0.890 / \textbf{2.698} & \textbf{1.2} / 0.9 & 1.036 / \textbf{1.427} & 10.996 / \textbf{11.049} & \textbf{0.301} / 0.150 & 0.148 / \textbf{0.181} & 4.067 / \textbf{9.133} & 56.523 / \textbf{60.108} \\
 & \rfive & \textbf{1.636} / 1.221 & \textbf{3.5} / 1.0 & \textbf{1.386} / 1.311 & 14.411 / \textbf{14.441} & \textbf{0.201} / 0.200 & 0.127 / \textbf{0.200} & 4.217 / \textbf{7.956} & 69.550 / \textbf{71.218} \\ \midrule
\multirow{5}{*}{\begin{tabular}[c]{@{}l@{}}Efficiency\\ (\trg$\uparrow$)\end{tabular}} & \rone & 20.433 / \textbf{21.024} & 0.2 / \textbf{0.4} & 2.002 / \textbf{2.249} & \textbf{8.482} / 8.477 & 0.249 / \textbf{0.350} & 0.142 / \textbf{0.183} & 4.550 / \textbf{9.461} & \textbf{60.415} / 53.401 \\
 & \rtwo & 21.585 / \textbf{23.024} & 0.1 / \textbf{0.2} & 1.345 / \textbf{2.444} & 9.279 / \textbf{9.507} & \textbf{0.299} / 0.249 & 0.127 / \textbf{0.222} & 4.744 / \textbf{11.417} & 70.531 / \textbf{71.965} \\
 & \rthree & 21.695 / \textbf{22.066} & \textbf{0.7} / 0.6 & 2.275 / \textbf{2.615} & 8.136 / \textbf{8.822} & \textbf{1.849} / 0.299 & 0.136 / \textbf{0.186} & 4.611 / \textbf{11.683} & 55.667 / \textbf{60.087} \\
 & \rfour & 23.958 / \textbf{24.606} & \textbf{1.2} / 0.8 & 1.036 / \textbf{1.506} & 10.996 / \textbf{11.016} & \textbf{0.301} / 0.150 & 0.148 / \textbf{0.190} & 4.067 / \textbf{7.717} & 56.523 / \textbf{64.172} \\
 & \rfive & 31.189 / \textbf{33.217} & 3.5 / \textbf{5.8} & 1.386 / \textbf{2.194} & 14.411 / \textbf{14.505} & 0.201 / \textbf{0.300} & 0.127 / \textbf{0.180} & 4.217 / \textbf{9.956} & 69.550 / \textbf{78.784} \\ \bottomrule
\end{tabular}

%% file: tables/tabRQ2.tex
\begin{tabular}{lllllllllllll}
\toprule
\multirow{2}{*}{\textbf{Metric}} & \multicolumn{4}{l}{\textbf{Collision Avoidance (R1)}} & \multicolumn{4}{l}{\textbf{Stability (R2)}} & \multicolumn{4}{l}{\textbf{Efficiency (R3)}} \\ \cmidrule(r){2-5} \cmidrule(r){6-9} \cmidrule(r){10-13}
 & \textit{\textbf{p}} & \textbf{\Atwelve} & \textbf{\rvsgb} & \textbf{\rvsgw} & \textit{\textbf{p}} & \textbf{\Atwelve} & \textbf{\rvsgb} & \textbf{\rvsgw} & \textit{\textbf{p}} & \textbf{\Atwelve} & \textbf{\rvsgb} & \textbf{\rvsgw} \\ \midrule
\dto & \textbf{$\bm{<}$0.01} & \textbf{0.238} & 0.776 & 0.656 & \textbf{$\bm{<}$0.01} & \textbf{0.253} & 0.496 & 0.494 & \textbf{$\bm{<}$0.01} & \textbf{0.242} & 0.315 & 0.375 \\
\jerk & $\geq$0.05 & 0.507 & 1.834 & 1.445 & $<$0.01 & 0.615 & 2.296 & 0.902 & $<$0.01 & 0.602 & 0.584 & 0.809 \\
\trg & $\geq$0.05 & 0.504 & 1.014 & 0.857 & $<$0.05 & 0.581 & 0.943 & 0.381 & $<$0.01 & 0.595 & 1.041 & 0.594 \\
\rpc & $<$0.05 & 0.579 & 5.579 & 2.734 & $<$0.01 & 0.582 & 1.150 & 0.392 & $<$0.01 & 0.599 & 1.777 & 1.611 \\
\chc & $<$0.05 & 0.568 & 0.921 & 0.693 & \textbf{$\bm{<}$0.01} & \textbf{0.710} & 0.625 & 0.232 & \textbf{$\bm{<}$0.01} & \textbf{0.797} & 0.737 & 0.294 \\
\pl & $\geq$0.05 & 0.547 & 0.160 & 0.097 & $\geq$0.05 & 0.546 & 0.323 & 0.065 & $\geq$0.05 & 0.560 & 0.151 & 0.053 \\
\tnm & $\geq$0.05 & 0.555 & 0.277 & 0.298 & $\geq$0.05 & 0.477 & 0.712 & 0.264 & $\geq$0.05 & 0.520 & 0.266 & 0.252 \\ \midrule
\ssd & / & / & 22.067 & 21.556 & / & / & 17.690 & 18.973 & / & / & 21.559 & 13.138 \\ \bottomrule
\end{tabular}